\documentclass[reprint,amsmath,amssymb,aps,]{revtex4-2}
\usepackage{graphicx}
\usepackage{subfigure}
\usepackage{placeins}
\usepackage{dcolumn}
\usepackage{bm}
\usepackage{cleveref}

\begin{document}
	
	\title{Estimation of Newtonian noise from KAGRA cooling system}

	\author{Rishabh Bajpai}
	\email{bajpai@post.kek.jp}
	\affiliation{The Graduate University for Advanced Studies, Department of Accelerator Science, School of High Energy Accelerator Science, High Energy Accelerator Research Organization (KEK) Tsukuba, Ibaraki, 305-0801, Japan}
	
	\author{Takayuki Tomaru}
	\affiliation{National Astronomical Observatory of Japan, 2-21-1 Osawa, Mitaka, Tokyo, 181-8588, Japan}
	\affiliation{The University of Tokyo, Department of Astronomy, Graduate School of Science, 2-21-1 Osawa, Mitaka, Tokyo, 181-8588, Japan}
	\affiliation{The Graduate University for Advanced Studies, 2-21-1 Osawa, Mitaka, Tokyo, 181-8588, Japan}
	\affiliation{High Energy Accelerator Research Organisation, 1-1 Oho, Tsukuba, Ibaraki, 305-0801, Japan}
	
	\author{Toshikazu Suzuki}
	\affiliation{Institute for Cosmic Ray Research (ICRR), The University of Tokyo, 5-1-5, Kashiwanoha, Kashiwa, Chiba, 277–8582, Japan}
	
	\author{Kazuhiro Yamamoto}
	\affiliation{Department of Physics, University of Toyama, Toyama, Toyama 930-8555, Japan}
	
	\author{Takafumi Ushiba}
	\affiliation{The University of Tokyo Institute for Cosmic Ray Research Kamioka Observatory, Higashimozumi 238, Kamioka, Hida, Gifu, 506-1205, Japan}
	
	\author{Tohru Honda}
	\affiliation{High Energy Accelerator Research Organisation, 1-1 Oho, Tsukuba, Ibaraki, 305-0801, Japan}
	
	\date{\today}% It is always \today, today,
	
	\begin{abstract}
		KAGRA is the first km-scale gravitational wave detector to be constructed underground and employ cryogenics to cool down it's test masses. 
		While the underground location provides a quiet site with low seismic noise, the cooling infrastructure is known to generate large mechanical vibrations due to cryocooler operation and structural resonances of the cryostat. 
		As cooling system components are relatively heavy and in close proximity to the test masses, oscillation of gravity force induced by their vibration, so-called Newtonian noise, could contaminate the detector sensitivity. 
		In this paper, we use the results from vibration analysis of the KAGRA cryostat to estimate cooling system Newtonian noise in the 1-100 Hz frequency band. 
		Our calculations show that, while this noise does not limit the current detector sensitivity or inspiral range, it will be an issue in the future when KAGRA improves its sensitivity.
		We conclude that KAGRA may need to implement Wiener filters to subtract this noise in the future.
	\end{abstract}
		
	\maketitle
	
	%\tableofcontents
	
	\section{\label{sec:1}Introduction}
	
Ground-based gravitational wave (GW) detectors are ultra-sensitive laser interferometers that detect displacements of the order of $10^{-19}$ m between mirrors several kilometers apart, making it essential to minimize environmental displacement noise.
Seismic noise (ground motion) is one of the fundamental noise sources for second-generation GW detectors like KAGRA \cite{1,2}, LIGO \cite{3} and Virgo \cite{4}. 
It couples to the Test Masses in two ways: (a) direct mechanical coupling and (b) Newtonian noise (NN). 
While, the mechanical coupling of seismic noise can be well attenuated by suspending the test masses from highly efficient vibration isolation systems \cite{5,6}, Newtonian noise cannot be shielded against. 

Newtonian noise  is caused by fluctuation of local gravitational fields caused by fluctuating mass distributions around test masses. 
Over the past few decades, several studies have been conducted to evaluate the NN from seismic fields, vibrating objects and atmospheric fields \cite{7,8,9,10,11}, but their contribution as noise sources are limited to the low frequency, outside the sensitivity band for second-generation detectors. 
However, Newtonian noise will be a serious issue for third-generation detectors like Einstein Telescope \cite{12} and Cosmic Explorer \cite{13}. 
Einstein Telescope aims to push the low-frequency sensitivity of ground-based GW detectors to its limit by increasing the arm length, using cryogenics to cool down the test masses and constructing detectors underground. 

The Large-scale Cryogenic Gravitational Wave Telescope (KAGRA), is a Dual Recycled Fabry-Perot Michelson Interferometer based GW detector located in Gifu, Japan. Compared to other second-generation detectors it has two unique features; first, the detector is located underground, which provides a quiet site (low seismic noise) and second, the mirrors are cooled down to 20 K reducing the thermal noises. 
It is these features that make KAGRA an ideal testbed for the development of the 3rd generation detectors. 

While some studies have been conducted at the KAGRA site to evaluate the seismic and atmospheric NN \cite{14} and underground water NN \cite{15}; NN contribution from the KAGRA Cooling system \cite{16} has not been evaluated. 
Vibration analysis of the cryostat at cryogenic temperature \cite{17} showed that vibration of the cooling system components surrounding the test masses is 2-3 orders larger than the seismic motion in the 1-100 Hz band due to structural resonances and cryocooler operation. 
As these components are relatively heavy and in close proximity to the TM, their NN contribution could contaminate the detector sensitivity.
A similar study for the LIGO vacuum chamber was conducted \cite{9} and concluded that NN from the chamber did not contaminate the detector sensitivity.

In this paper, we present methods, considerations, calculations and results of Newtonian noise from KAGRA Cooling system.
\section{Theory}
\subsection{Derivation of Expression for Newtonian Noise} \label{sec:nnestimation}
\begin{figure} [b]
	\centering
	\subfigure[]{\includegraphics[width=0.15\textwidth]{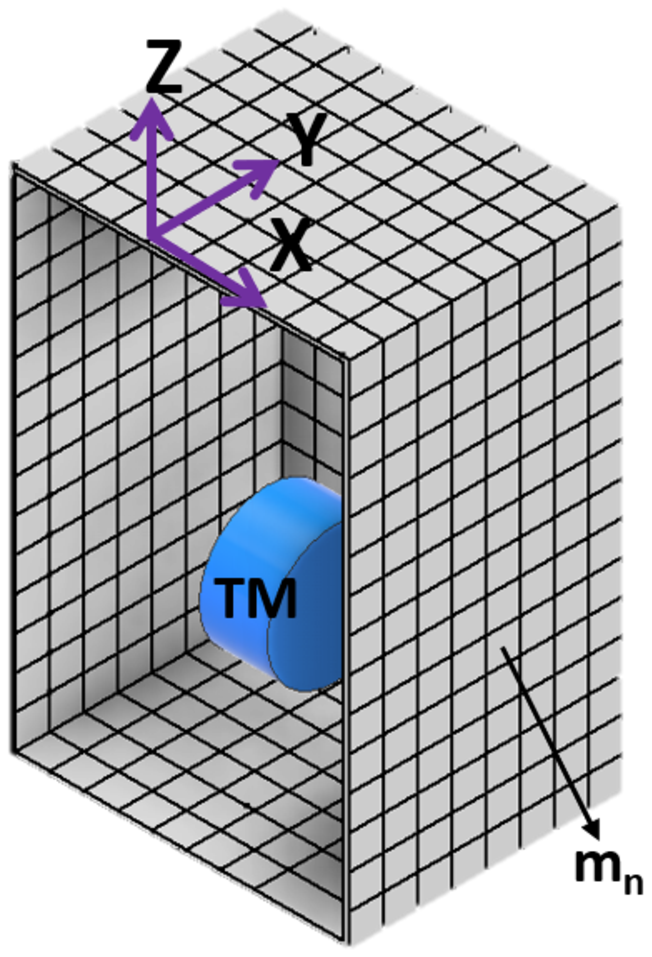}\label{fig:1a}}
	\subfigure[]{\includegraphics[width=0.25\textwidth]{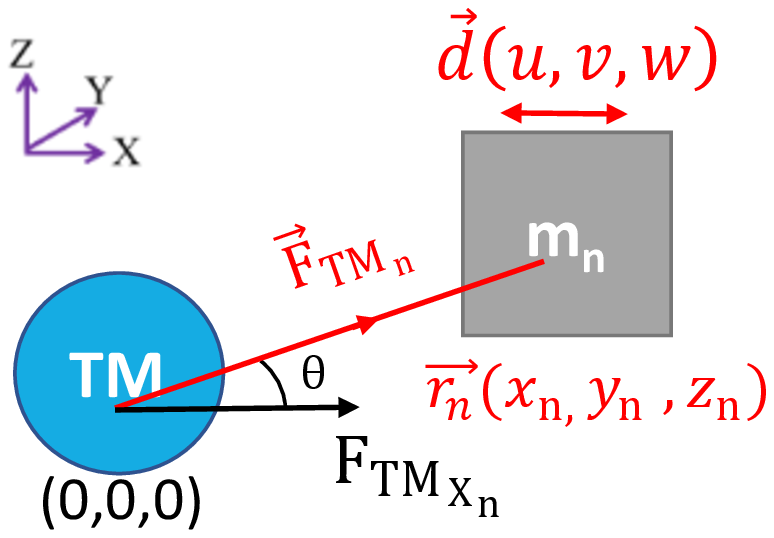}\label{fig:1b}}
	\caption{(a) A simple hollow cuboid cooling system-Test Mass (TM) configuration is considered to derive Newtonian noise expression. The TM is located at the origin, and the cuboid is split into N elements each weighing $m_n$ and denoted by the grey squares. (b) Newtonian force on TM due to a single element on cooling system at position $\protect\overrightarrow{r}(x_{n},y_{n},z_{n})$ is $\protect\overrightarrow{F}_{\mathrm{TM}}$. The mass is being displaced(/vibrating) by a small amount $\protect\overrightarrow d(u,v,w)$, such that$|r|\protect\gg|d|$. The coordinate system considered is shown with	purple arrows and X-axis is parallel to the optical-axis.} 
	\label{fig:1}
\end{figure}
To derive the expression for Newtonian Noise, we consider a simple hollow cuboid-shaped cooling system with the test mass (TM) (Sapphire Mirror, of mass $M$=23 kg) as a point mass located at the origin, as shown in \cref{fig:1a}.
The entire cuboid cooling system is considered as a mass distribution made of N point-masses (finite elements), each weighing $m_{n}$ and denoted grey by squares.
For a mass $m_{n}$ at a position $\overrightarrow{r_n}(x_n,y_n,z_n)$ on the cooling system, the Newtonian force on TM is:
\begin{equation}\label{1}
\overrightarrow{F}_{\mathrm{TM_n}}=GMm_n\frac{\overrightarrow{r}}{|\overrightarrow{r}|^{3}}
\end{equation}

where $G$ is the gravitational constant.
\Cref{fig:1b} shows this system.
Now, the $X$-component of this force (along the optical axis of the interferometer) is:

\begin{equation}\label{2}
F_{\mathrm{TM_{X_{n}}}}=GMm_{n}\frac{x_{n}}{(x_{n}^{2}+y_{n}^{2}+z_{n}^{2})^{\frac{3}{2}}}
\end{equation}

The gradient of this expression is:

\begin{equation}\label{3}
\begin{split}
{\nabla{F_{\mathrm{TM_{X_{n}}}}}=\frac{\partial F_{\mathrm{TM_{X_{n}}}}}{\partial x_{n}}\hat{i}+\frac{\partial F_{\mathrm{TM_{X_{n}}}}}{\partial y_{n}}\hat{j}+\frac{\partial F_{\mathrm{TM_{X_{n}}}}}{\partial z_{n}}\hat{k}}\\
{\resizebox{0.4\textwidth}{!}{$=GMm_{n} \left[\frac{(-2x_{n}^{2}+y_{n}^{2}+z_{n}^{2})\hat{i}-3x_{n}y_{n}\hat{j}-3z_{n}x_{n}\hat{k}}{(x_{n}^{2}+y_{n}^{2}+z_{n}^{2})^{\frac{5}{2}}}\right]$}}
\end{split}
\end{equation}
where $\hat{i}$, $\hat{j}$ and $\hat{k}$ are unit vector along $X$, $Y$ and $Z$ axis, respectively.

If mass $m_n$ moves/vibrates by a small amount $\overrightarrow{d}$;
\begin{equation}\label{4}
\overrightarrow{d}=u(t)\hat{i}+v(t)\hat{j}+w(t)\hat{k}
\end{equation}
The change in force on test mass along the $X$-axis due to displacement $\overrightarrow{d}$ is the scalar product $\nabla{F_{\mathrm{TM_{X_{n}}}}}.\overrightarrow{d}$, which gives us:
\begin{equation}\label{5}
\resizebox{0.43\textwidth}{!}{$\delta{F_{\mathrm{TM_{X_{n}}}}}(t)=GMm_{n} \left[\frac{(-2x_{n}^{2}+y_{n}^{2}+z_{n}^{2})u(t)-3x_{n}y_nv(t)-3z_{n}x_nw(t)}{(x_{n}^{2}+y_{n}^{2}+z_{n}^{2})^{\frac{5}{2}}}\right]$}
\end{equation}
Ignoring the internal resonances of the cooling system, it is assumed that each element moves with the same displacement. 
So the change in force on TM due to the entire cooling system (CS) will be;
%\begin{equation}\label{6}
%\begin{split}
%{\delta{F_{\mathrm{TM_{X_{CS}}}}}(t)=GM\Bigg[ \sum_{n=1}^{N}m_{n}\left(\frac{-2x_{n}^{2}+y_{n}^{2}+z_{n}^{2}}{(x_{n}^{2}+y_{n}^{2}+z_{n}^{2})^{\frac{5}{2}}}\right)u(t)}\\
%{+\sum_{n=1}^{N}m_{n}\left(\frac{-3x_{n}y_{n}}{(x_{n}^{2}+y_{n}^{2}+z_{n}^{2})^{\frac{5}{2}}}\right)v(t)}
%\\{+\sum_{n=1}^{N}m_{n}\left(\frac{-3z_{n}x_{n}}{(x_{n}^{2}+y_{n}^{2}+z_{n}^{2})^{\frac{5}{2}}}\right)w(t)\Bigg]}
%\\{=GM[Au(t)+Bv(t)+Cw(t)]}
%\end{split}
%\end{equation}
\begin{equation}\label{6}
{\delta{F_{\mathrm{TM_{X_{CS}}}}}(t)=GM[Au(t)+Bv(t)+Cw(t)]}
\end{equation}
where,

$A=\sum_{n=1}^{N}m_{n}\left[\frac{-2x_{n}^{2}+y_{n}^{2}+z_{n}^{2}}{(x_{n}^{2}+y_{n}^{2}+z_{n}^{2})^{\frac{5}{2}}}\right]$

$B=\sum_{n=1}^{N}m_{n}\left[\frac{-3x_{n}y_{n}}{(x_{n}^{2}+y_{n}^{2}+z_{n}^{2})^{\frac{5}{2}}}\right]$ 

$C=\sum_{n=1}^{N}m_{n}\left[\frac{-3z_{n}x_{n}}{(x_{n}^{2}+y_{n}^{2}+z_{n}^{2})^{\frac{5}{2}}}\right]$

are constants. Now, the Fourier transform of \cref{6} is:
\begin{equation}\label{7}
\delta F_{\mathrm{TM_{X_{CS}}}}(\omega)=GM[Au(\omega)+Bv(\omega)+Cw(\omega)]
\end{equation}
where, $\omega$ is the angular frequency. 

For power spectral density of an ergodic signal $P(t)$ defined as: $\big \langle P(\omega)P^{*}(\omega')\big \rangle=S_{P}(\omega)\delta(\omega-\omega')$, the amplitude spectral density of $\delta{F_{\mathrm{TM_{X_{CS}}}}}(t)$, $\sqrt{S_{F_{\mathrm{TM_{X_{CS}}}}}(\omega)}$ will be:
\begin{equation}\label{8}
\resizebox{0.42\textwidth}{!}{$\sqrt{S_{F_{\mathrm{TM_{X_{CS}}}}}(\omega)}=GM\sqrt{A^{2}S_u(\omega)+B^{2}S_v(\omega)+C^{2}S_w(\omega)}$}
\end{equation}
where, $S_u(\omega),\,S_v(\omega)\, \mathrm{and}\,S_w(\omega)$ are power spectral densities of the cooling system vibration along $X$, $Y$ and $Z$ axis and it is assumed they have no correlation between them.

Now \cref{8} gives us the amplitude spectral density of force fluctuation, dividing it by $1/M\omega^{2}$ gives us displacement spectral density for one test mass. Since, similar vibration coupling with no correlation for all four test masses is expected, a factor of $\sqrt{4}$ is multiplied. Finally the dividing the expression with detector arm-length, $L$ (=3000 m) gives us the corresponding strain amplitude of cooling system Newtonian noise as:
\begin{equation}\label{9}
\resizebox{0.43\textwidth}{!}{$\sqrt{S_{h}(\omega)}=\frac{\sqrt{4}G}{\omega^{2}L}\sqrt{A^{2}S_u(\omega)+B^{2}S_v(\omega)+C^{2}S_w(\omega)}\equiv H$}
\end{equation}
%where
%$1/\omega^{2}$ converts the acceleration to displacement of the test mass,
%$\sqrt{4}$ is a multiplicative factor for 4 test masses, and
%$L$ = 3000 m is the arm-length.
\subsection{Considerations for KAGRA Cooling System}
\begin{figure}[t]
	\centering
	\includegraphics[width=0.5\textwidth]{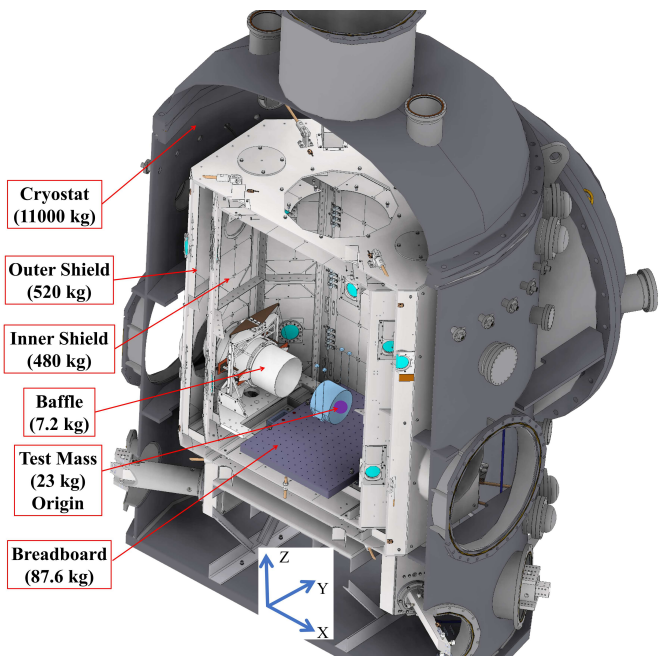}
	\caption[Coordinate system and summary of components considered in Newtonian Noise Calculation]{Coordinate system and summary of components considered in Newtonian noise calculation. Cross-section of KAGRA cryostat showing the components (chamber, baffle,breadboard, inner shield and outer shield) considered for Newtonian noise calculation. The blue arrows denote the coordinate system. The test mass is considered as a point mass, located at origin as denoted by the purple dot.}
	\label{fig:2}
\end{figure}
In KAGRA the 23-kg sapphire test masses are suspended from a nine-stage vibration isolation system and cooled down inside a double radiation shield cryostat \cite{16}.
\Cref{fig:2} shows the cross-section of the KAGRA cryostat and components (chamber, baffle, breadboard, inner and outer radiation shield) considered for Newtonian noise calculation are denoted with red boxes.
The mirror is considered to be a point-mass at the origin. The coordinate system for the NN calculation is denoted by the blue arrows as:
\begin{itemize}
	\item \textit{X-Axis} is the optical axis of the interferometer.
	\item \textit{Y-Axis:} is the horizontal axis perpendicular to the optical axis.
	\item \textit{Z-Axis:} is the vertical axis
\end{itemize}
\begin{table}[h!]
%	\vspace{-6mm}
	\caption{The values of the summations A, B and C of various components for the generated mesh.}
	\label{tab:0}
	\begin{center}
		\begin{tabular}{|c|c|c|c|}
			\hline
			\textbf{Component} & \textbf{A} & \textbf{B} & \textbf{C}\\
			\hline
			Breadboard & 550.69 & $-5.4\times10^{-16}$ & $-1.2\times10^{-14}$\\
			\hline
			Baffle & -23.4 & -6.97 & -0.11\\
			\hline
			Inner Shield & 124.15 & -0.28 & -4.64\\
			\hline
			Outer Shield & 97.52 & 0.53 & -1.95\\
			\hline
			Cryostat & -298.34 & -6.22 & -0.03\\
			\hline
		\end{tabular}
	\end{center}
\end{table}
The expression derived in \cref{9} gives the Newtonian noise strain for simple mass distribution-test mass configuration, for the KAGRA cooling system the mass distribution is much more complex.
Newtonian noise is evaluated by generating a mesh with N elements in Ansys Mechanical Enterprise (software based on finite-element method) and solving it for \cref{9}.
\Cref{tab:0} shows the value of A,B, C as evaluated from the mesh generated for each component.

Since A is dominant value, we only consider the effect of vibration along the X-axis (optical axis) to estimate the Newtonian noise coupling of each component. The \cref{9} is then reduced to:

\begin{equation}\label{10}
\begin{split}
{H_{\mathrm{NN}}=\frac{\sqrt{4}G}{\omega^{2}L} \times \sqrt{S_u(\omega)}\times A}\\
{=\frac{K}{\omega^{2}}\times u_{\mathrm{component}} \sum_{n=1}^{N}S_{n_{\mathrm{component}}}}
\end{split}
\end{equation}
where,

$K=\frac{\sqrt{4}G}{L}=4.44 \times 10^{-14}\, \mathrm{m}^2\mathrm{kg}^{-1}\mathrm{s}^{-2}$; 

$S_{n_{\mathrm{component}}}= \left[m_n\left( \frac{(-2x_n^{2}+y_n^{2}+z_n^{2})}{(x_n^{2}+y_n^{2}+z_n^{2})^{\frac{5}{2}}}\right) \right]\,\mathrm{kg}/\mathrm{m}^{-3}$

$u_{\mathrm{component}}=\sqrt{S_u(\omega)}$, is displacement spectral density of vibration (in $\mathrm{m/\sqrt{Hz}}$) of the component under consideration.
\section{Calculation} \label{sec:3}
\subsection{Breadboard}
\begin{figure}[t]
	\centering
	\subfigure[] {\includegraphics[width=0.25\textwidth]{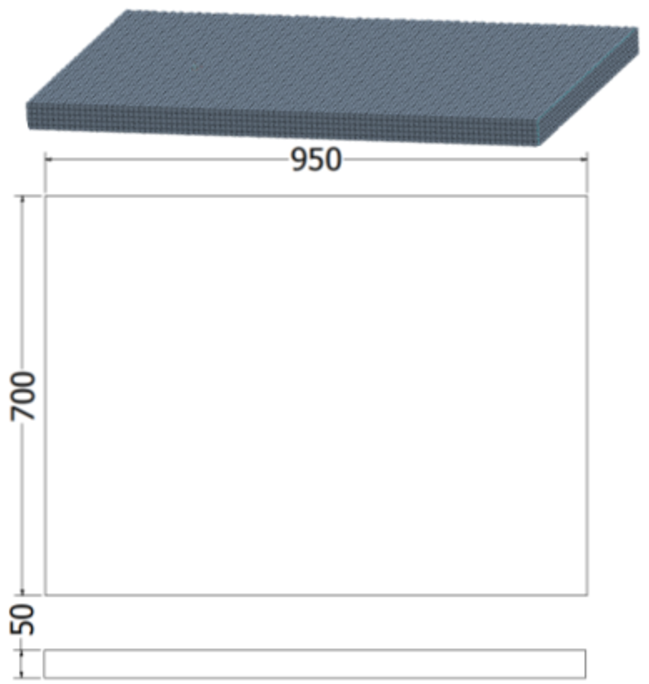}\label{fig:3a}}	
	\subfigure[] {\includegraphics[width=0.45\textwidth]{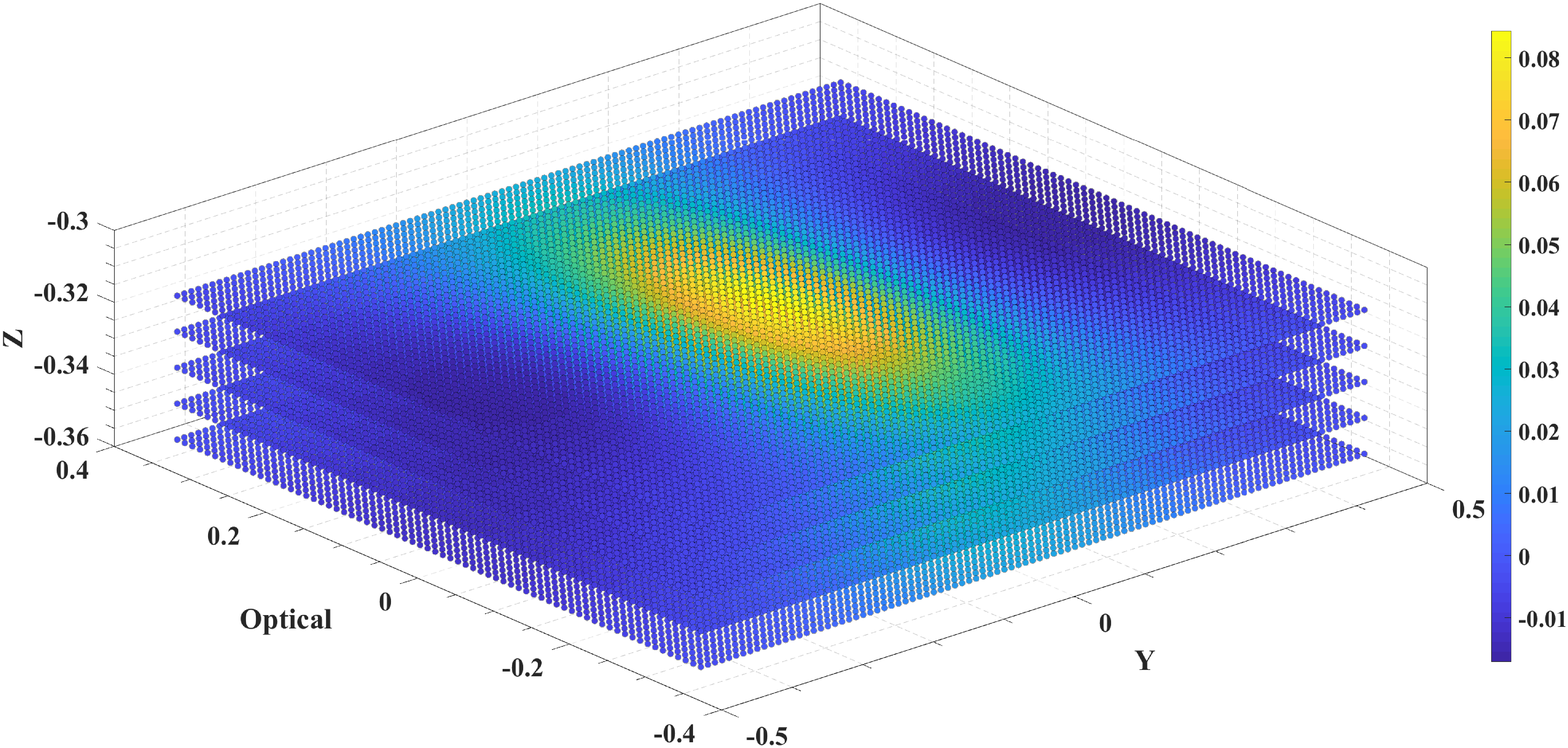}\label{fig:3b}}	
	\caption{(a) Mesh generated and 2-D drawing for the breadboard. The cuboid is divided into 33,250 elements each weighing 2.634 grams. (b) 3-D scatter plot where each dot represents an element n from breadboard mesh while the color is value $S_{n_{\mathrm{breadboard}}}$  of that element in \cref{11}}
\end{figure}

The breadboard is an aluminum cuboid plate of dimension $0.7\times0.95\times0.05$ m and mass of 87.6 kg at a distance of 0.335 m from the TM. 
The board is bolted to the bottom of the inner shield and is used to fix the earthquake stop frame for the cryogenic payload \cite{18}. 
The generated mesh had elements size of 0.01 m splitting cuboid into 33,250 elements, each weighing 2.634 g. \Cref{fig:3a} shows the generated mesh and 2-D drawing of breadboard. Based on the generated mesh \cref{10} becomes:
\begin{equation}\label{11}
H_{\mathrm{breadboard}}=\frac{K}{\omega^{2}}\times u_{\mathrm{breadboard}}\sum_{n=1}^{33250}S_{n_{\mathrm{breadboard}}}
\end{equation}

\begin{figure} [h] 
	\centering
	\subfigure[] {\includegraphics[width=0.32\textwidth]{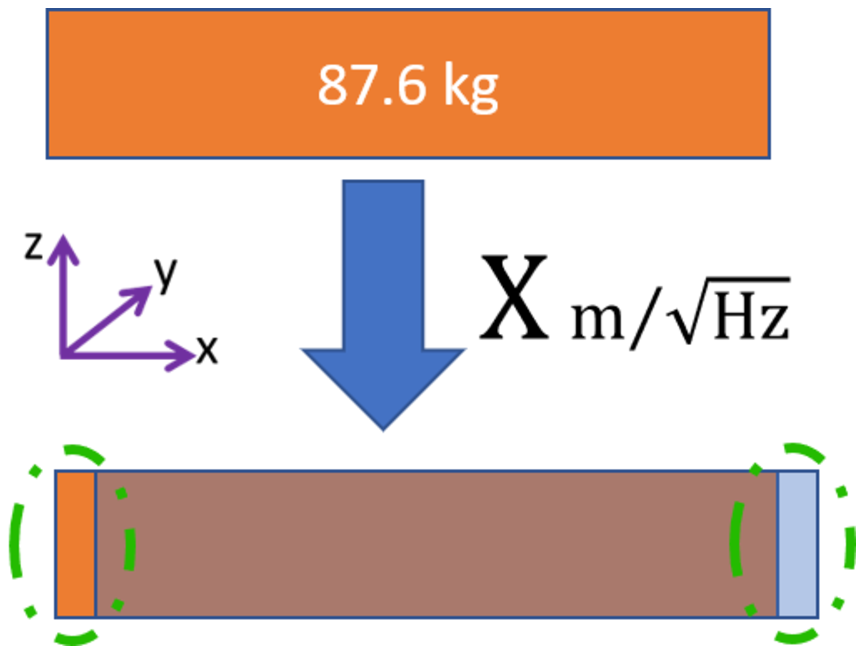}\label{fig:4a}}	
	\subfigure[] {\includegraphics[width=0.38\textwidth]{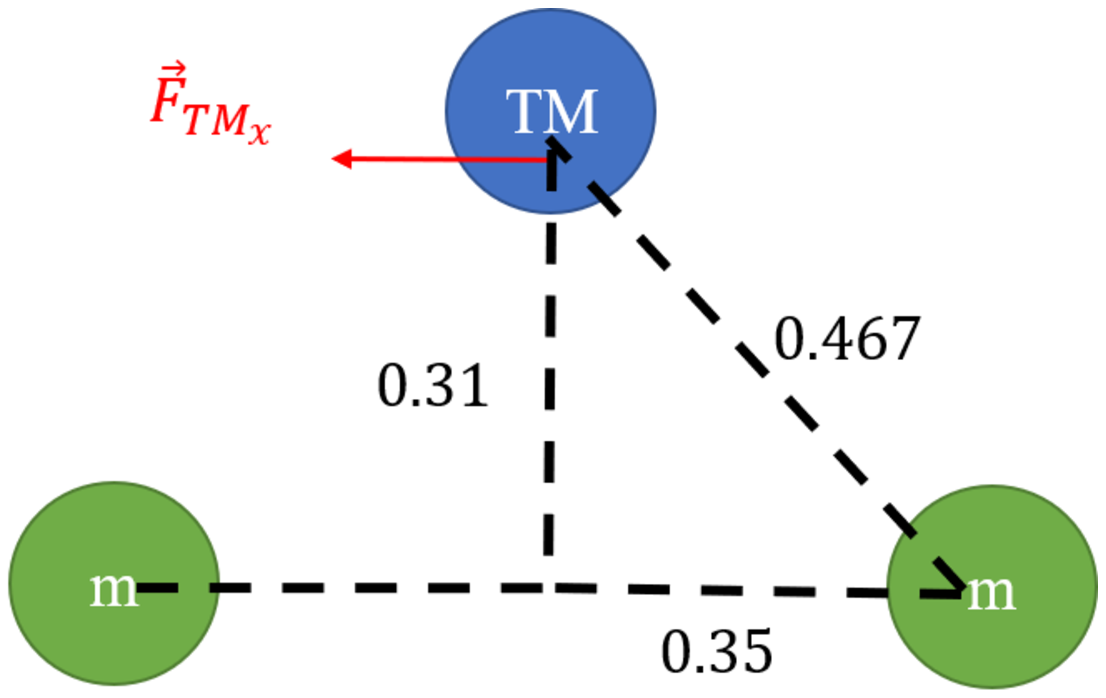}\label{fig:4b}}
	\caption[Simplified breadboard-Test Mass system for direct calculation]{(a) Front view of the breadboard, being displaced X m/$\surd$Hz. The dotted green circle represent mass that causes gravity fluctuations. (b) A simple point mass representation of \cref{fig:4a} showing relative position of Test Mass and breadboard point masses m.}
	\label{fig:4}
\end{figure}
%\begin{figure} [h!]
%	\centering
%	\subfigure[]{\includegraphics[width=0.3\textwidth]{./figures/breadboardintegral.eps}\label{fig:breadboardintegral}}
%	\subfigure[] {\includegraphics[width=0.3\textwidth]{./figures/breadboardreduced.eps}\label{fig:breadboardreduced}}
%	\caption[Breadboard integral solution.]{(a) Simplified optical breadboard below the test mass for NN calculation. (b) A simple rectangle (green) considered at center if the breadboard to simplify the NN calculation. \label{fig:breadboardsimple}}
%\end{figure}
The value of summation  $\sum_{n=1}^{33250}S_{n_{\mathrm{breadboard}}}$ is 550.697 $\mathrm{kg}/\mathrm{m}^{-3}$, substituting the  values in \cref{11} we get,
\begin{equation}\label{12}
H_{\mathrm{breadboard}}=2.295\times 10^{-11}\times \frac{u_{\mathrm{breadboard}}}{\omega^{2}}
\end{equation}
where, $u_{\mathrm{breadboard}}$ is the vibration (/displacement) measured \cite{17} at 12 K.

A simplified system, shown in \cref{fig:4} was considered to confirm the simulation result.
\Cref{fig:4a} shows the front view of the breadboard. 
Now, if the displacement (vibration) spectra of the breadboard is  $X$ m/$\surd$Hz at some frequency $f$. 
The force acting on the TM will only be due to the small mass, represented by green dotted circles (in \cref{fig:4a}) displaced at the ends of the breadboard. 
A simplified representation of this is shown in \cref{fig:4b}, from which the force on TM along X-axis can be calculated as:
\begin{equation}\label{13}
\overrightarrow{F}_{TM_{X_{\mathrm{breadboard}}}}=2.G.M.m.\frac{0.35}{0.467^{3}}
\end{equation}
where $m$ is $87.6X$ kg at some frequency $f$ and $M$ is mass of Test Mass
From \cref{13} the expression for strain can be derived as:
\begin{equation}\label{14}
H_{\mathrm{{breadboard}}}=\frac{\sqrt{4}}{\omega^{2}.L}\times 2\times G\times 87.6X\times 3.436
\end{equation}
where, $X=u_\mathrm{{bottom}}$ is the vibration measured by cryogenic accelerometer. 
\begin{equation}\label{15}
H_{\mathrm{{breadboard}}}=2.676\times 10^{-11}\times \frac{u_\mathrm{{bottom}}}{\omega^{2}}
\end{equation}
%A simplified system, shown in \cref{fig:breadboardsimple} was considered to confirm the simulation result.
%The cuboid breadboard (in \cref{fig:breadboardintegral}) was reduced to a rectangle of dimension $0.7\times0.95$ m at the center of cuboid with the mass of 87.6 kg evenly distributed across the surface. 
%The distance of this rectangle from the TM is 0.31 m (=$z$). 
%A small mass $dm$ on this rectangle is $M_{\mathbf{unit}}$ $dxdy$, where:
%\begin{equation} \label{13}
%M_{\mathbf{unit}}=\frac{\mathrm{Mass\, of\, the\, breadboard}}{\mathrm{Surface\, area\, of\, the\, rectangle}}=\frac{87.6}{0.7\times0.95}=139.36
%\end{equation}
%From \cref{fig:breadboardreduced,13} the NN strain in \cref{10} can be written in integral form as:
%\begin{equation}\label{14}
%\begin{split}
%{H_{{\mathrm{breadboard}}}=\frac{\sqrt{4}GM_{\mathbf{unit}}u_\mathrm{{breadboard}}}{\omega^{2}L}}\\
%{\times \int_{-0.475}^{0.475}\int_{-0.35}^{0.35}\frac{(-2x^{2}+y^{2}+z^{2})}{(x^{2}+y^{2}+z^{2})^{\frac{5}{2}}} dx\,dy}
%\end{split}
%\end{equation}

%\begin{equation}\label{15}
%H_{{\mathrm{breadboard}}}=2.828\times 10^{-11}\times \frac{u_\mathrm{{bottom}}}{\omega^{2}}
%\end{equation}
Comparing the \cref{12,15} it is clear that results NN coupling from breadboard, calculated by computer simulation and by hand are comparable and there is not significant error.
\subsection{Baffle}

The wide-angle baffle (WAB) is a hollow cylinder suspended 10 mm in front of the HR side of each TM to prevent the back-scattered light from recoupling to the main beam as it is coated in SolBlack.
Cylinder is 570 mm long with inner diameter of 254 mm, 4 mm thickness, is made of aluminum and weighs about 2 kg. 
The total suspended mass is about 7.2 kg and the natural frequency of the longitudinal mode is about 0.84 Hz. 
A detailed review of WAB can be found at \cite{19,20}.

\begin{figure}[t]
	\centering
	\subfigure[] {\includegraphics[width=0.3\textwidth]{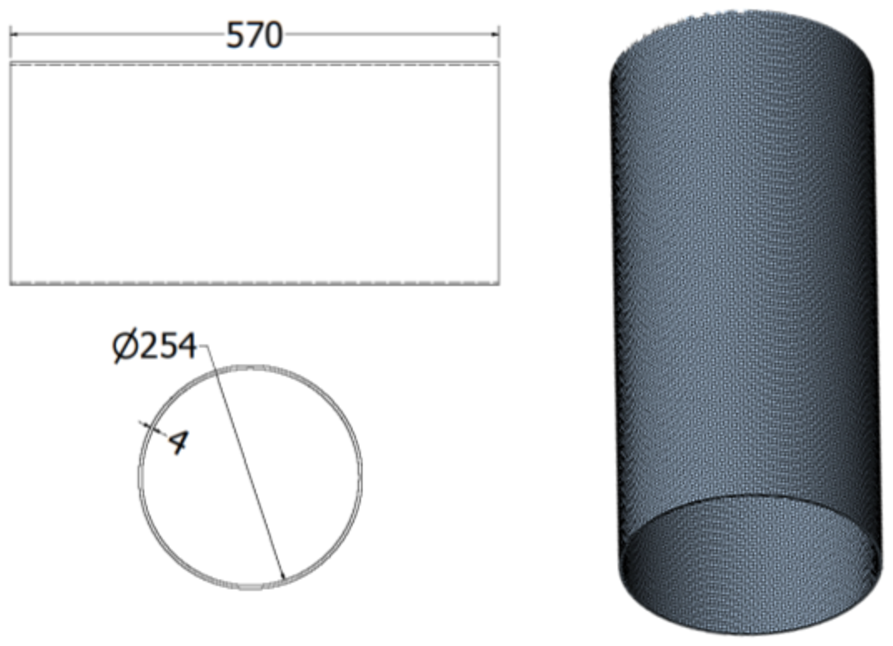}\label{fig:5a}}	
	\subfigure[] {\includegraphics[width=0.45\textwidth]{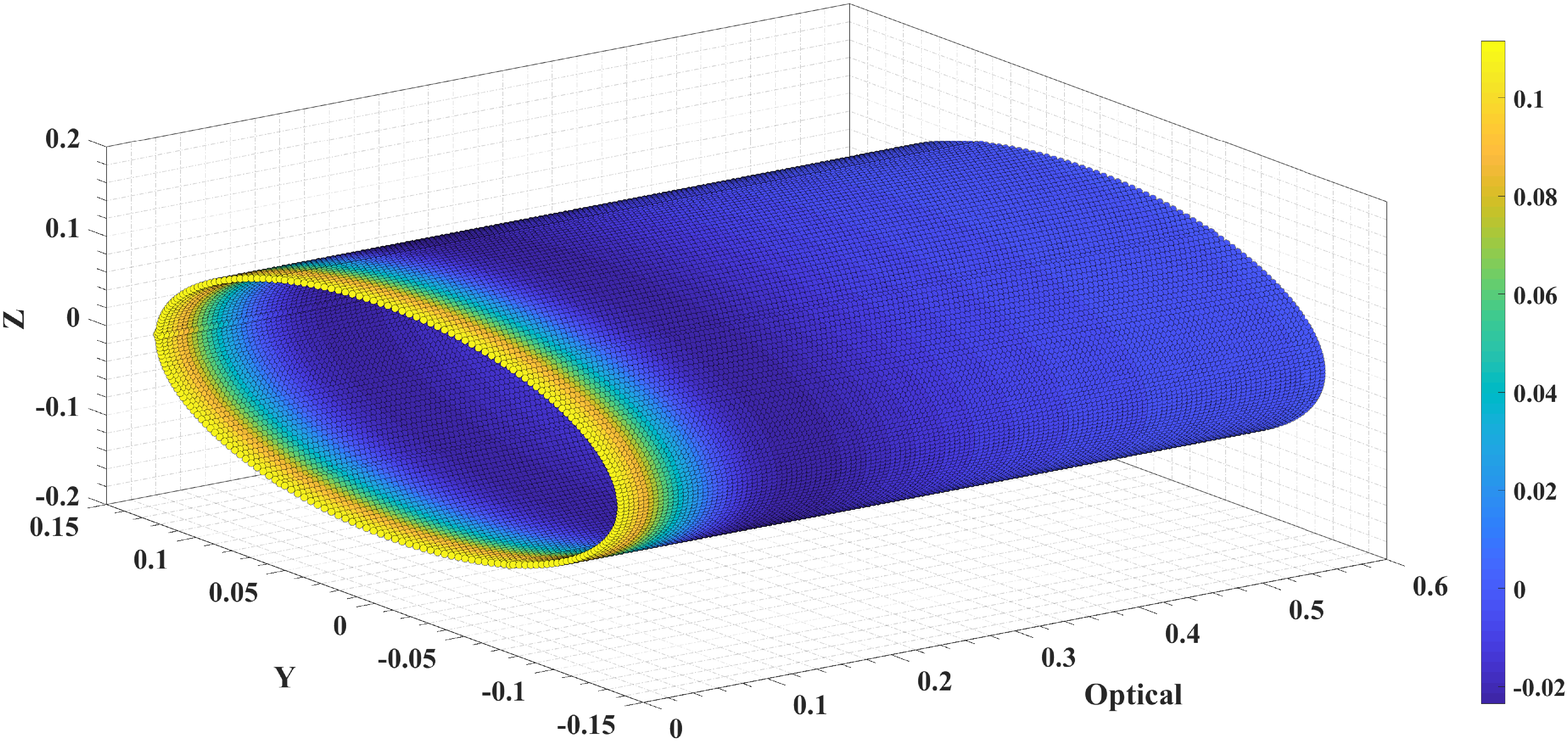}\label{fig:5b}}	
	\caption{(a) Mesh generated for baffle. The cylinder is divided into 29,601 elements each weighing 0.24 grams. (b) 3-D scatter plot where each dot represents an element n from baffle mesh while the color is value $S_{n_{\mathrm{baffle}}}$  of that element in \cref{16}}
\end{figure}

Note that for the calculations, entire 7.2 kg mass was assumed to be evenly distributed across the cylinder.
The generated mesh had elements size of 0.004 m splitting the cylinder into 29,601 elements, each weighing 0.24 g. \Cref{fig:5a} shows the generated mesh and 2-D drawing of the baffle. 
Based on the generated mesh \cref{10} becomes:
\begin{equation}\label{16}
H_{\mathrm{baffle}}=\frac{K}{\omega^{2}}\times u_{\mathrm{baffle}}\sum_{n=1}^{29601}S_{n_{\mathrm{baffle}}}
\end{equation}

The value of summation $\sum_{n=1}^{29601}S_{n_{\mathrm{baffle}}}$ is $-24.75$ $\mathrm{kg}/\mathrm{m}^{-3}$, substituting the  values in \cref{16} we get,
\begin{equation}\label{17}
H_{\mathrm{baffle}}=-9.76\times 10^{-13}\times \frac{u_{\mathrm{baffle}}}{\omega^{2}}
\end{equation}
where, $u_{\mathrm{baffle}}=u_{\mathrm{breadboard}}\times$ Transfer function of baffle suspension

The baffle suspension is rigidly bolted to the breadboard and since the longitudinal mode of the suspended baffle is 0.84 Hz, the transfer function is: $\frac{0.84^2}{f^2}$ for $f>0.84$ Hz, where, $f$ is the frequency. Therefore, \cref{17} becomes:
\begin{equation}\label{18}
H_{\mathrm{baffle}}=-6.88\times 10^{-13}\times \frac{u_{\mathrm{breadboard}}}{\omega^{2}f^{2}}
\end{equation}
\subsection{Thermal Radiation Shield}
Test masses are cooled down inside a double thermal radiation shield, which are octagonal prism structures with a combined weight of  $\sim$ 1000 kg. 
These shields, made of Aluminum (Al1070) are called inner/8K and outer/80K shield, weighing $\sim$ 480 kg and $\sim$ 520 kg respectively. As can be seen in \cref{fig:2}, shield structure is relatively complex including multiple components (like support beams, viewports, screws etc.), to simplify the calculation we considered a simple octagonal prism with same dimensions as the shields. 
The generated mesh and dimension of each shield can be found in \cref{fig:6a,fig:7a}.

\subsubsection{8 K Shield}
\begin{figure}[b]
	\centering
	\subfigure[] {\includegraphics[width=0.3\textwidth]{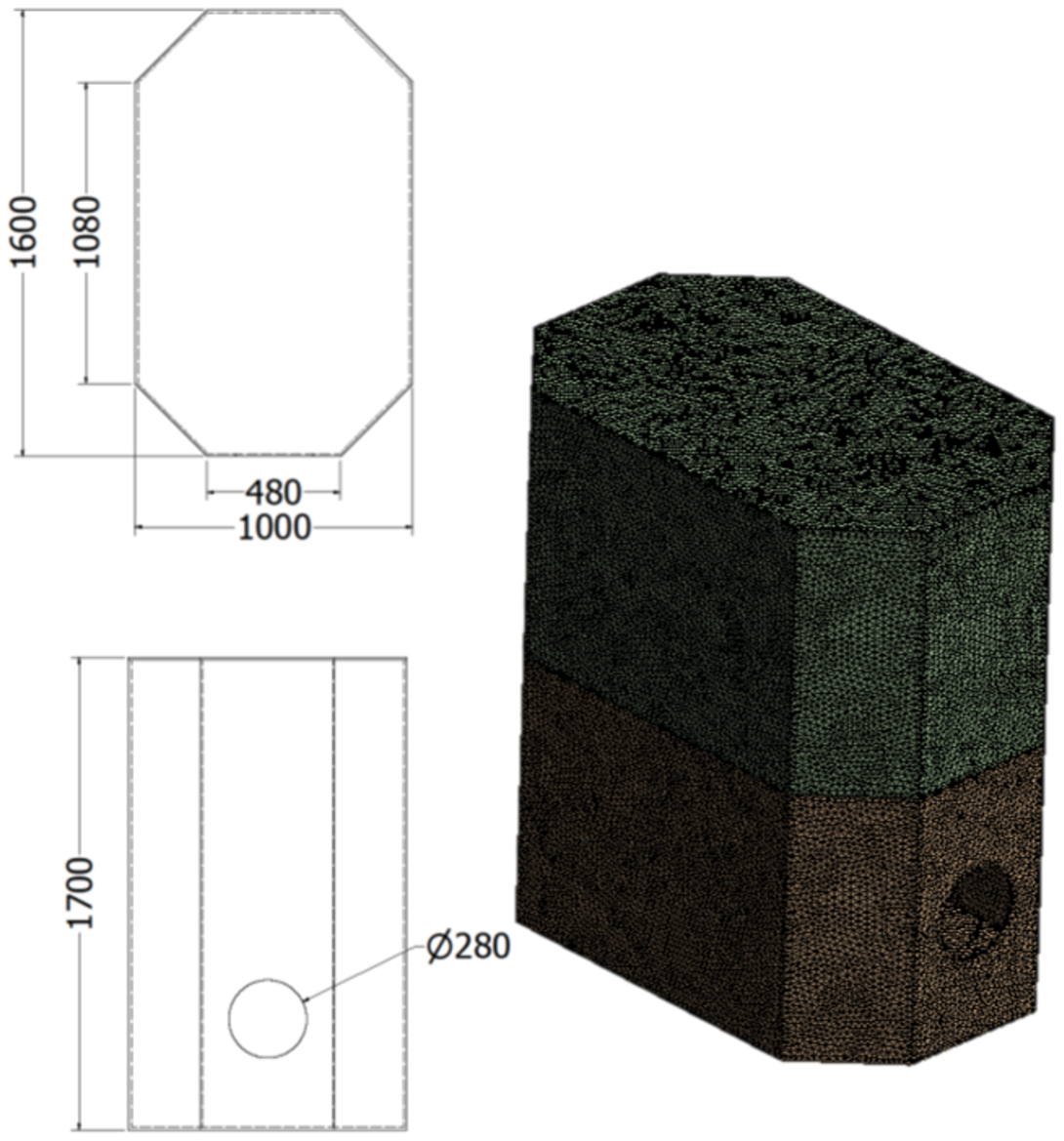}\label{fig:6a}}	
	\subfigure[] {\includegraphics[width=0.3\textwidth]{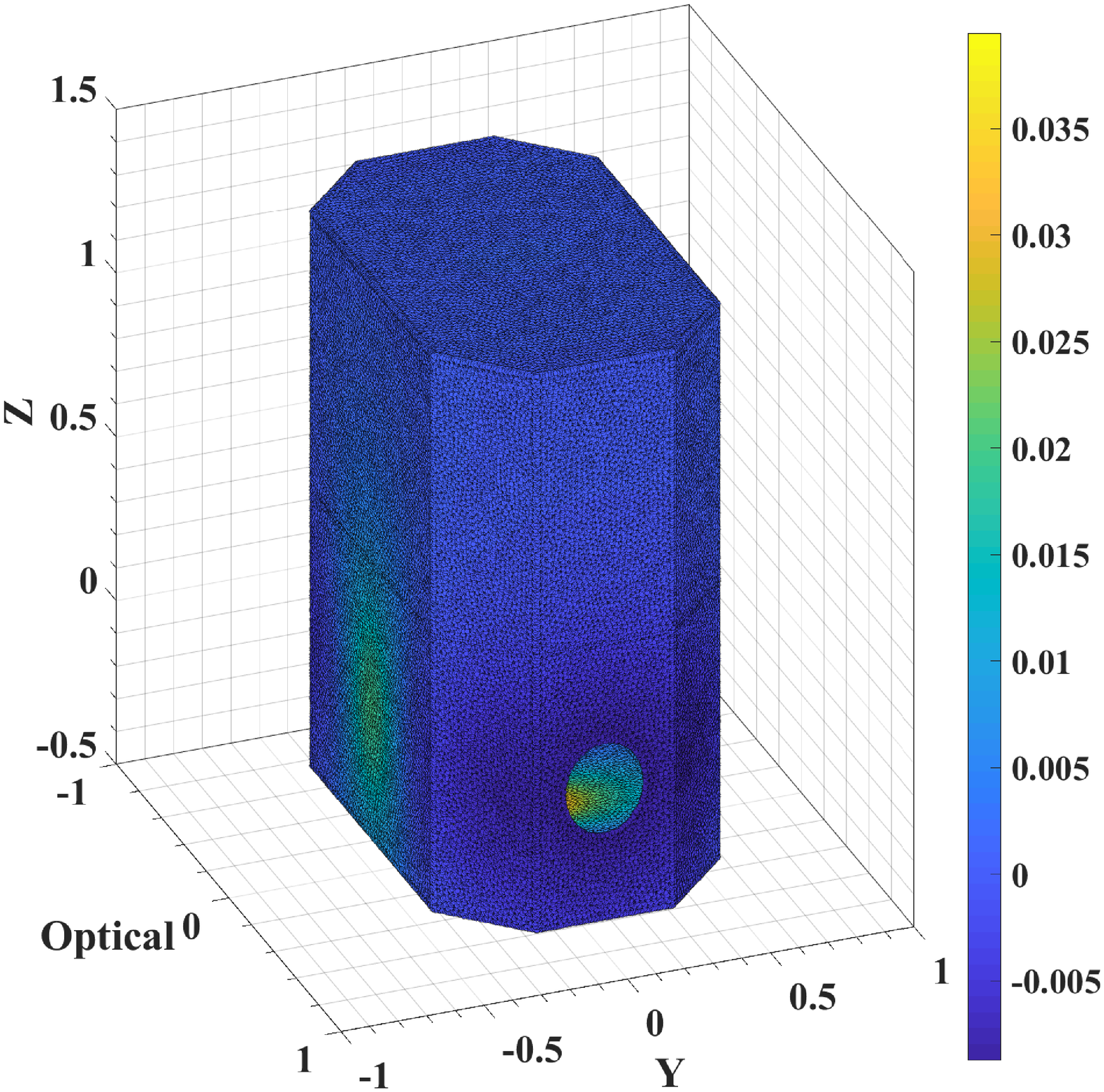}\label{fig:6b}}	
	\caption{(a) Mesh generated for inner shield with 200,935 elements each weighing 2.4 g. 2-D drawing showing top and front view of 8K shield. (b) 3-D scatter plot where each dot represents an element n from inner shield mesh while the color is value $S_{n_{\mathrm{8K}}}$  of element in \cref{19}}
\end{figure}
\Cref{fig:6a} shows the generated mesh for a simplified inner (8K) shield, based on which the expression for \cref{10} becomes:
\begin{equation}\label{19}
H_\mathrm{{8K}}=\frac{K}{\omega^{2}}\times u_\mathrm{{8K}}\sum_{n=1}^{200935}S_{n_\mathrm{{8K}}}
\end{equation}
In \cref{19} $u_\mathrm{{8K}}$ is the vibration (/displacement) of inner shield; due internal resonances of chamber and shield the magnitude of vibration  will increase as we move along the Z-axis. 
Vibration of elements on top ($u_\mathrm{{8K_{top}}}$) and bottom ($u_\mathrm{{8K_{bottom}}}$) surfaces will be constant while that of the walls will be a function of z ($u_\mathrm{{8K_{walls}}}(z)$). 

To precisely estimate the NN coupling we need to evaluate the displacement of wall elements for each resonance mode, as the computing resources were limited we split the shield into two equal halves and assume that the top and bottom section move with $u_\mathrm{{8K_{top}}}$ and $u_\mathrm{{8K_{bottom}}}$ respectively.  
From this \cref{19} can be simplified to:
\begin{equation}\label{20}
\begin{split}
{H_\mathrm{{8K}}=\frac{K}{\omega^{2}}\times \Bigg[ u_\mathrm{{8K_{bottom}}}\sum_{n=1}^{99882}S_{n_\mathrm{{8K_{bottom}}}}}\\
{+ u_\mathrm{{8K_{top}}}\sum_{n=99883}^{200935}S_{n_\mathrm{{8K_{top}}}}\Bigg]}
\end{split}
\end{equation}
The value of summation $\sum_{n=1}^{99882}S_{n_\mathrm{{8K_{bottom}}}}$ and $\sum_{n=99883}^{200935}S_{n_\mathrm{{8K_{top}}}}$ is 59.5 $\mathrm{kg}/\mathrm{m}^{-3}$ and 6.466 $\mathrm{kg}/\mathrm{m}^{-3}$ respectively, substituting these values in \cref{20} we get,
\begin{equation}\label{21}
H_\mathrm{{8K}}=2.48\times10^{-12}\times\frac{u_\mathrm{{8K_{bottom}}}}{\omega^{2}}+2.69\times10^{-12}\times\frac{u_\mathrm{{8K_{top}}}}{\omega^{2}}
\end{equation}

\subsubsection{80 K Shield}
\begin{figure}[t]
	\centering
	\subfigure[] {\includegraphics[width=0.3\textwidth]{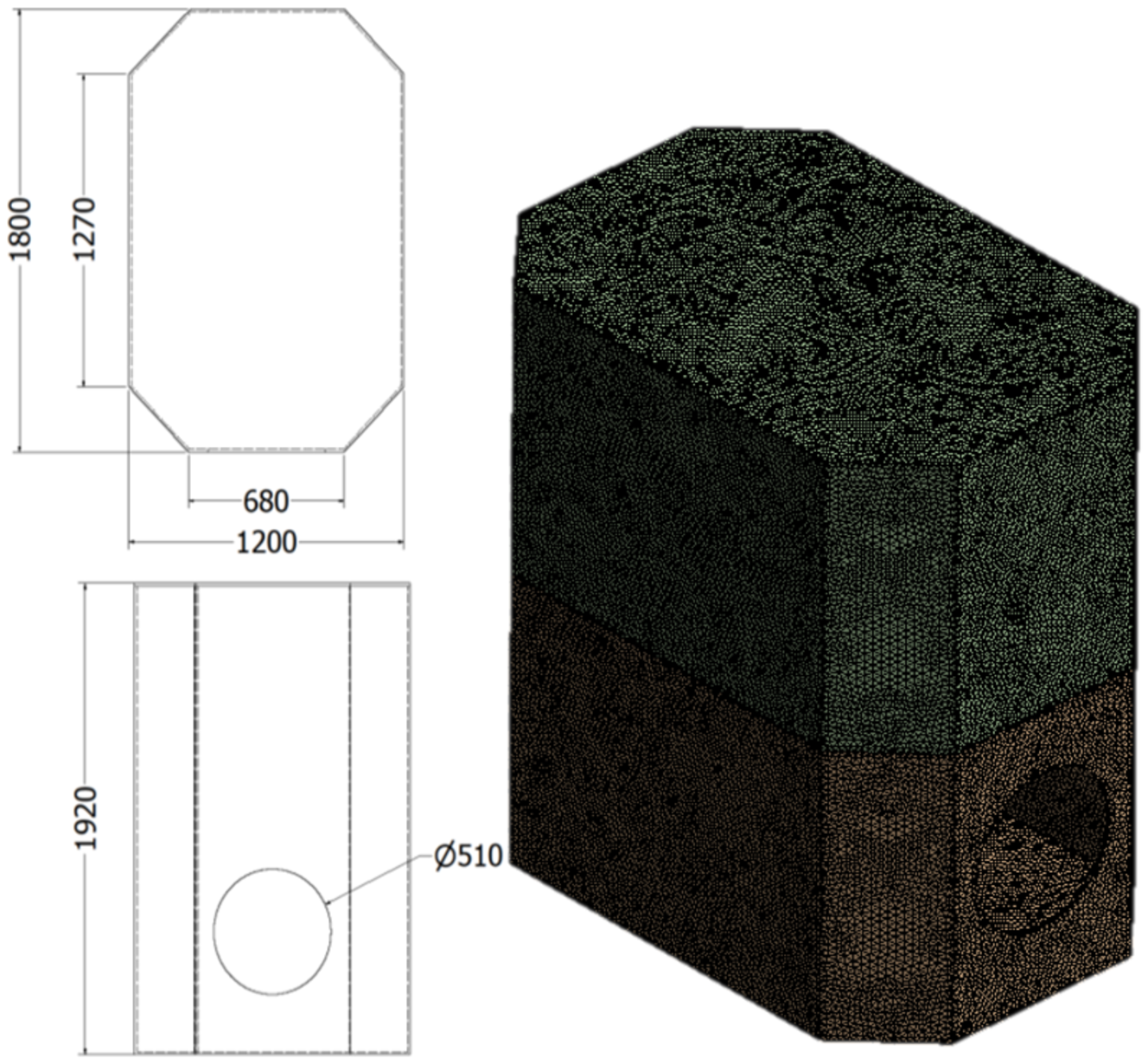}\label{fig:7a}}	
	\subfigure[] {\includegraphics[width=0.3\textwidth]{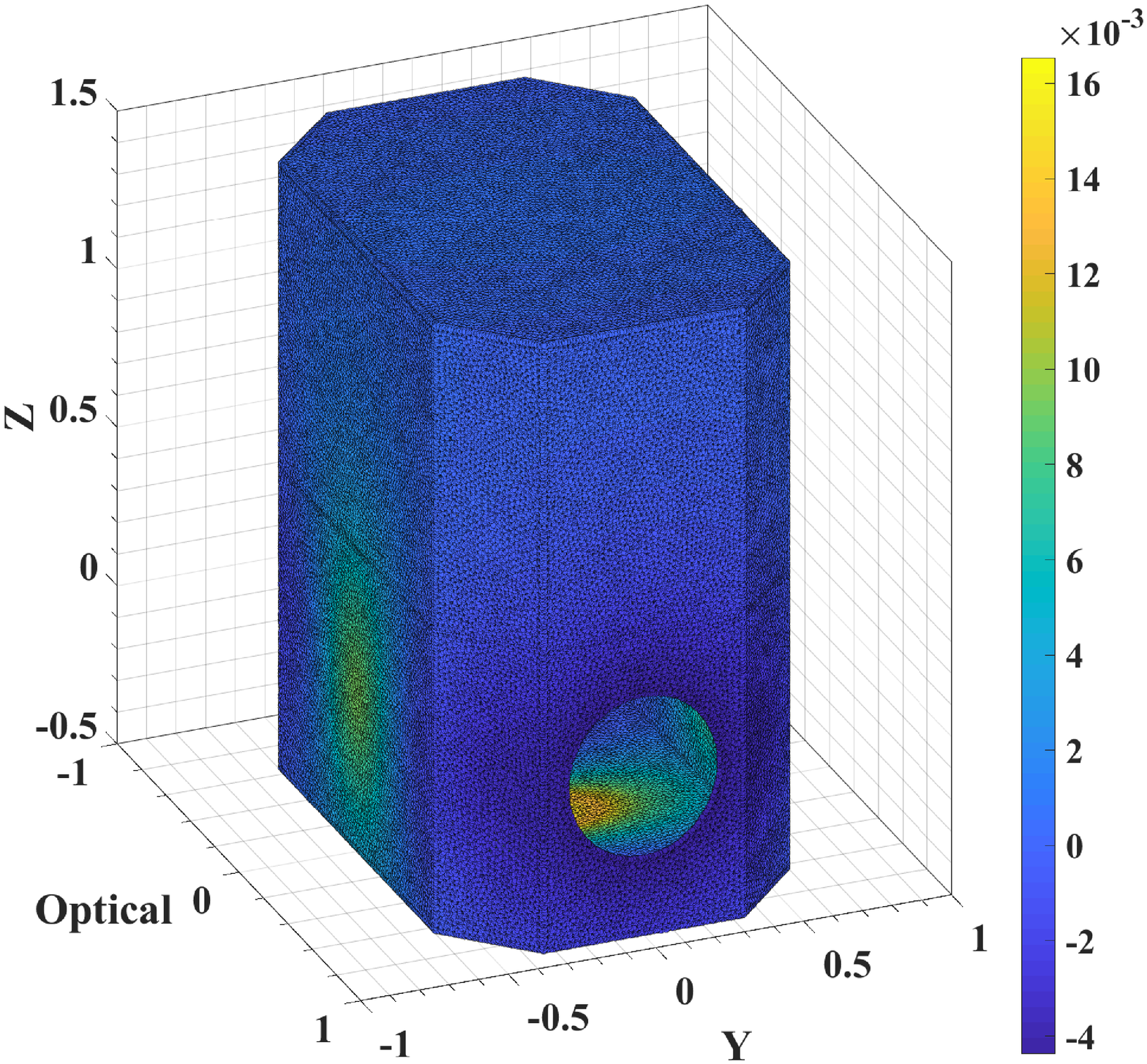}\label{fig:7b}}	
	\caption{(a) Mesh generated for outer shield with 262,889 elements each weighing 1.97 g. 2-D drawing showing top and front view of 80K shields. (b) 3-D scatter plot where each dot represents an element n from outer shield mesh while the color is value $S_{n_\mathrm{{80K}}}$  of that element in \cref{22}.}
\end{figure}
\Cref{fig:7a} shows the generated mesh for a simplified outer (80K) shield, based on which the expression for \cref{10} becomes:
\begin{equation}\label{22}
H_\mathrm{{80K}}=\frac{K}{\omega^{2}}\times u_\mathrm{{80K}}\sum_{n=1}^{262899}S_{n_\mathrm{{80K}}}
\end{equation}
For calculating the NN noise from 80K shield we make the same considerations as the 8K shield. 
As it was difficult to practically measure the 80 K shield vibration, we assume it to be same as that of the 8 K shield. 
80K shield was also split into two equal halves; the generated mesh had 262,899 elements each weighing 1.97 gm. 
Based on the generated mesh, \cref{22} becomes,
\begin{equation}\label{23}
H_\mathrm{{80K}}=2.34\times10^{-12}\times\frac{u_\mathrm{{80K_{bottom}}}}{\omega^{2}}+1.73\times10^{-12}\times\frac{u_\mathrm{{80K_{top}}}}{\omega^{2}}
\end{equation} 
where, $u_\mathrm{{80K_{bottom}}}=u_\mathrm{{8K_{bottom}}}$ and $u_\mathrm{{80K_{top}}}=u_\mathrm{{8K_{top}}}$ is assumed

\subsection{Cryostat Chamber}
\begin{figure}[t]
	\centering
	\subfigure[] {\includegraphics[width=0.28\textwidth]{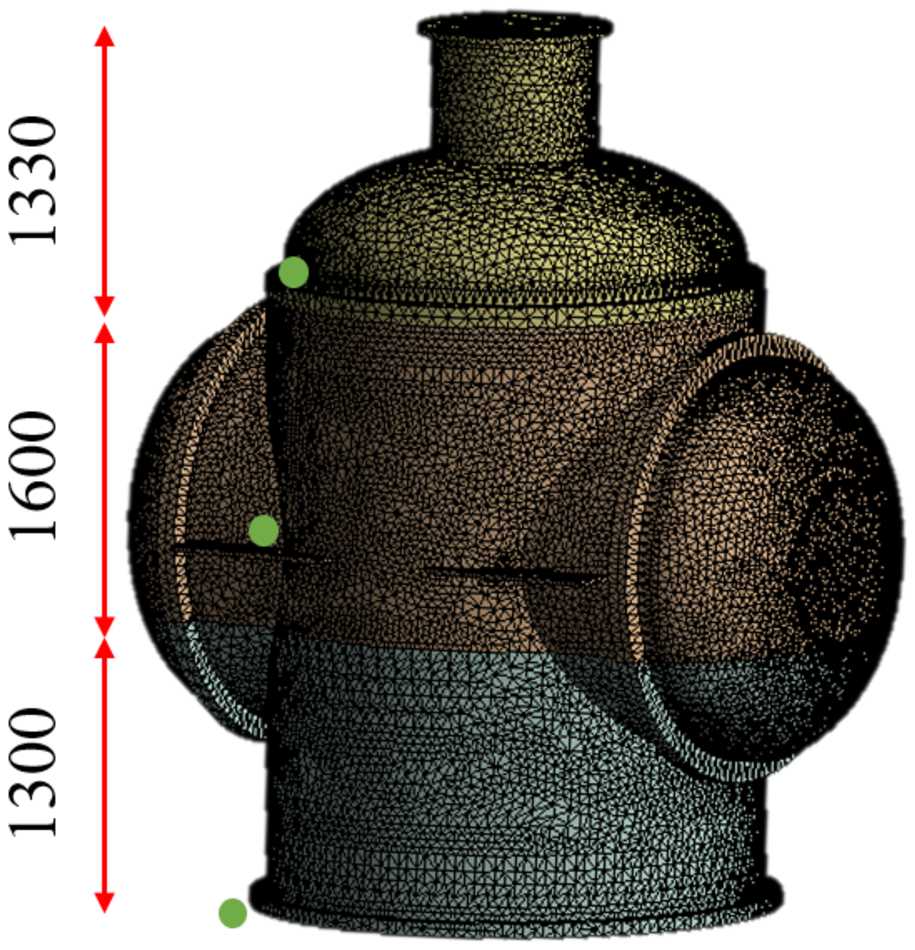}\label{fig:8a}}	
	\subfigure[] {\includegraphics[width=0.34\textwidth]{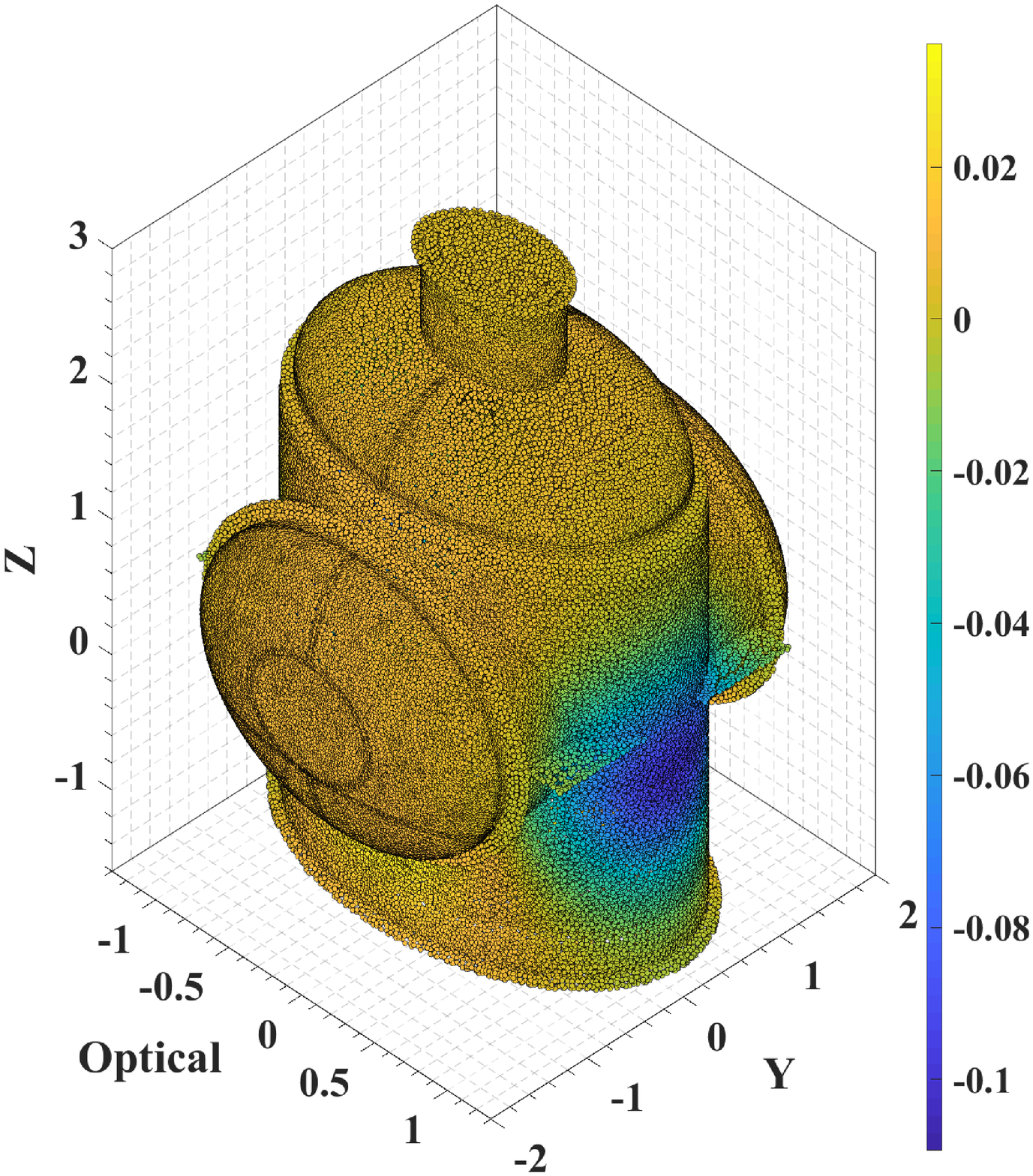}\label{fig:8b}}	
	\caption{(a) Mesh generated for cryostat with 124,795 elements. Cryostat was split into three parts and the green dots represent the position where vibrations $u_\mathrm{{top}}$, $u_\mathrm{{middle}}$ and $u_\mathrm{{bottom}}$ were measured. (b) 3-D scatter plot where each dot represents an element n from cryostat mesh while the color is value $S_{n_\mathrm{{cryostat}}}$  of that element in \cref{24}}
\end{figure}
The cryostat chamber (Height:4.33 m, Outer Diameter: 2.3 m) is a stainless steel structure (SUS-304) weighing about 11,000 kg. The structure itself has various flanges which we ignore in our calculation and consider a simple structure, shown in \cref{fig:8a} along with the mesh. Based on the generated mesh the expression for \cref{10} becomes: 
\begin{equation}\label{24}
H_\mathrm{{cryostat}}=\frac{K}{\omega^{2}}\times u_\mathrm{{cryostat}}\sum_{n=1}^{124795}S_{n_\mathrm{{cryostat}}}
\end{equation}
To evaluate NN from cryostat, we divide it into three sections: top, middle and bottom with vibrations $u_\mathrm{{top}}$, $u_\mathrm{{middle}}$ and $u_\mathrm{{bottom}}$, respectively. 
The expression in \cref{24} becomes:

\begin{equation}\label{25}
\begin{split}
{H_\mathrm{{cryostat}}=\frac{K}{\omega^{2}}\times \Bigg[ u_\mathrm{{top}}\sum_{n=1}^{18908}S_{n_\mathrm{{top}}}}\\
{ + u_\mathrm{{middle}}\sum_{n=60254}^{124795}S_{n_\mathrm{{middle}}}
+ u_\mathrm{{bottom}}\sum_{n=18909}^{60253}S_{n_\mathrm{{bottom}}}\Bigg] }
\end{split}
\end{equation}
The value of summation $\sum_{n=1}^{18908}S_{n_\mathrm{{top}}}$, $\sum_{n=60254}^{124795}S_{n_\mathrm{{middle}}}$ and $\sum_{n=18909}^{60253}S_{n_\mathrm{{bottom}}}$ are 101.99 $\mathrm{kg}/\mathrm{m}^{-3}$, -317.9 $\mathrm{kg}/\mathrm{m}^{-3}$ and -82.43 $\mathrm{kg}/\mathrm{m}^{-3}$ respectively, substituting these values in \cref{25} we get:
\begin{equation}\label{26}
\begin{split}
{H_\mathrm{{cryostat}}=6.8\times10^{-9}\times\frac{u_\mathrm{{top}}}{\omega^{2}}}\\
{-2.12\times10^{-8}\times\frac{u_\mathrm{{middle}}}{\omega^{2}}-5.49\times10^{-9}\times\frac{u_\mathrm{{bottom}}}{\omega^{2}}}
\end{split}	
\end{equation} 
\subsection{Comparison}
\begin{figure}[t]
	\centering
	\includegraphics[width=0.5\textwidth]{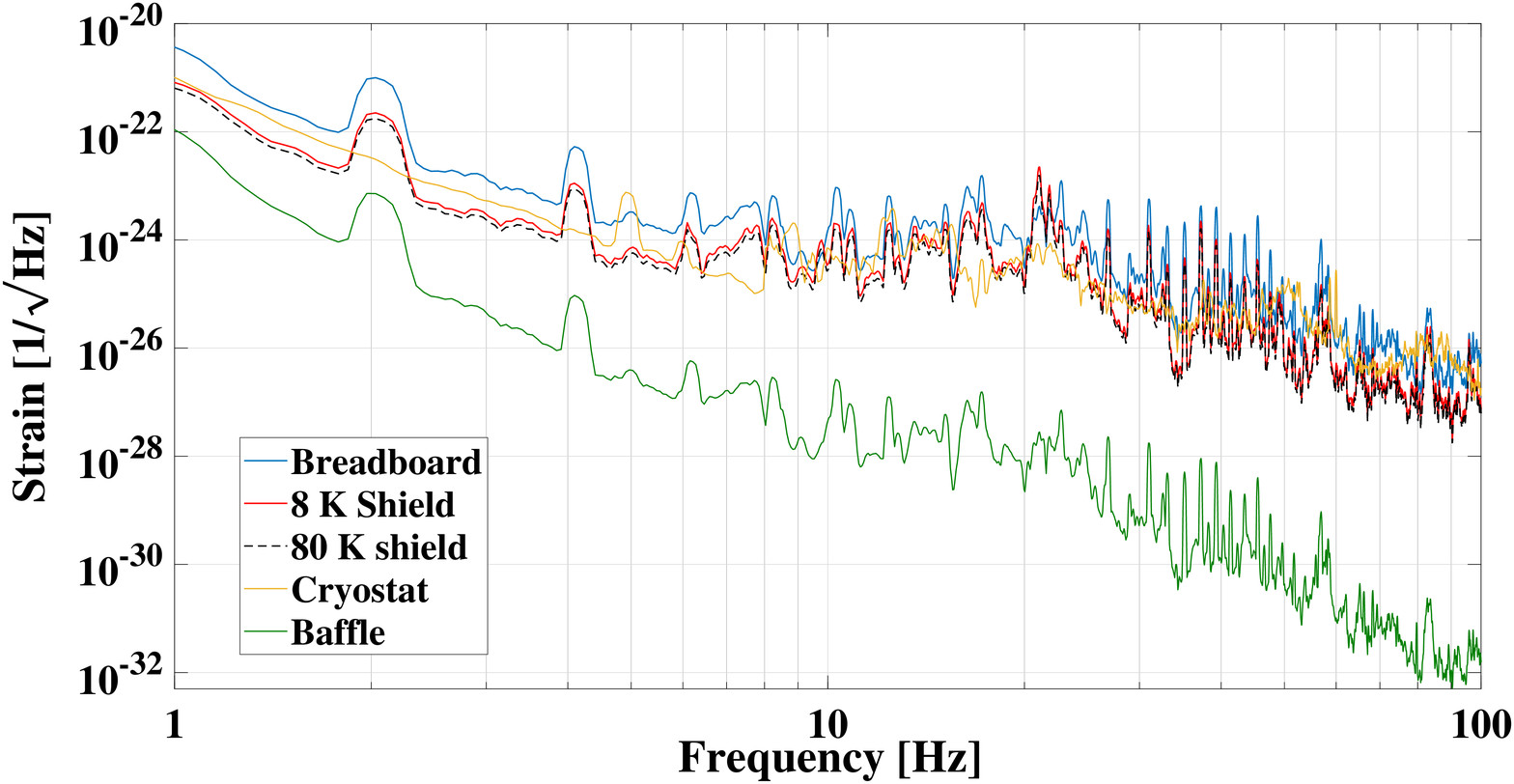}
	\caption{Comparison of Newtonian Noise strain spectral density from each component, the largest contribution comes from breadboard followed by radiation shields, cryostat and baffle. Note that the force on Test Mass due breadboard/radiation shields and cryostat/baffle are in opposite directions.}
	\label{fig:9}
\end{figure}

Substituting the measured vibration of each component (\cref{fig:11}) in \cref{12,18,21,23,26} we get the Newtonian noise strain for breadboard, baffle, inner shield, outer shield and cryostat respectively as shown in \cref{fig:9}. 
Note that the displacement of TM due to breadboard/radiation shields and cryostat/baffle will be in opposite directions.

Breadboard NN dominates the entire 1-100 Hz spectra except for 21-23 Hz peaks where radiation shields NN dominates, because these are resonance mode of the shields. 
Even though cryostat chamber in 10 times heavier than radiation shield, it's NN contribution is lower/comparable over the entire spectra due to symmetry and as it is further away from the TM. While, baffle is the closet object to TM it's NN contribution is the lowest because of vibration attenuation from the baffle suspension and relatively low mass.

\section{Impact on sensitivity}
\begin{figure}[t]
	\centering
	\subfigure[] {\includegraphics[width=0.5\textwidth]{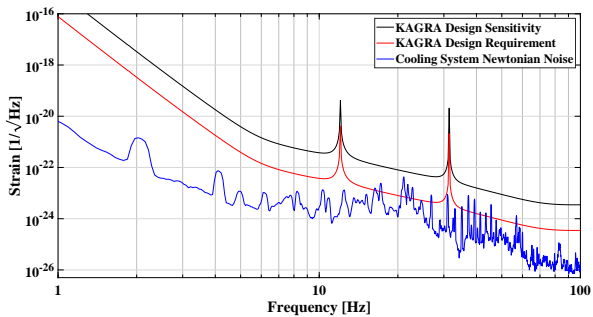}\label{fig:10a}}
	\subfigure[] {\includegraphics[width=0.5\textwidth]{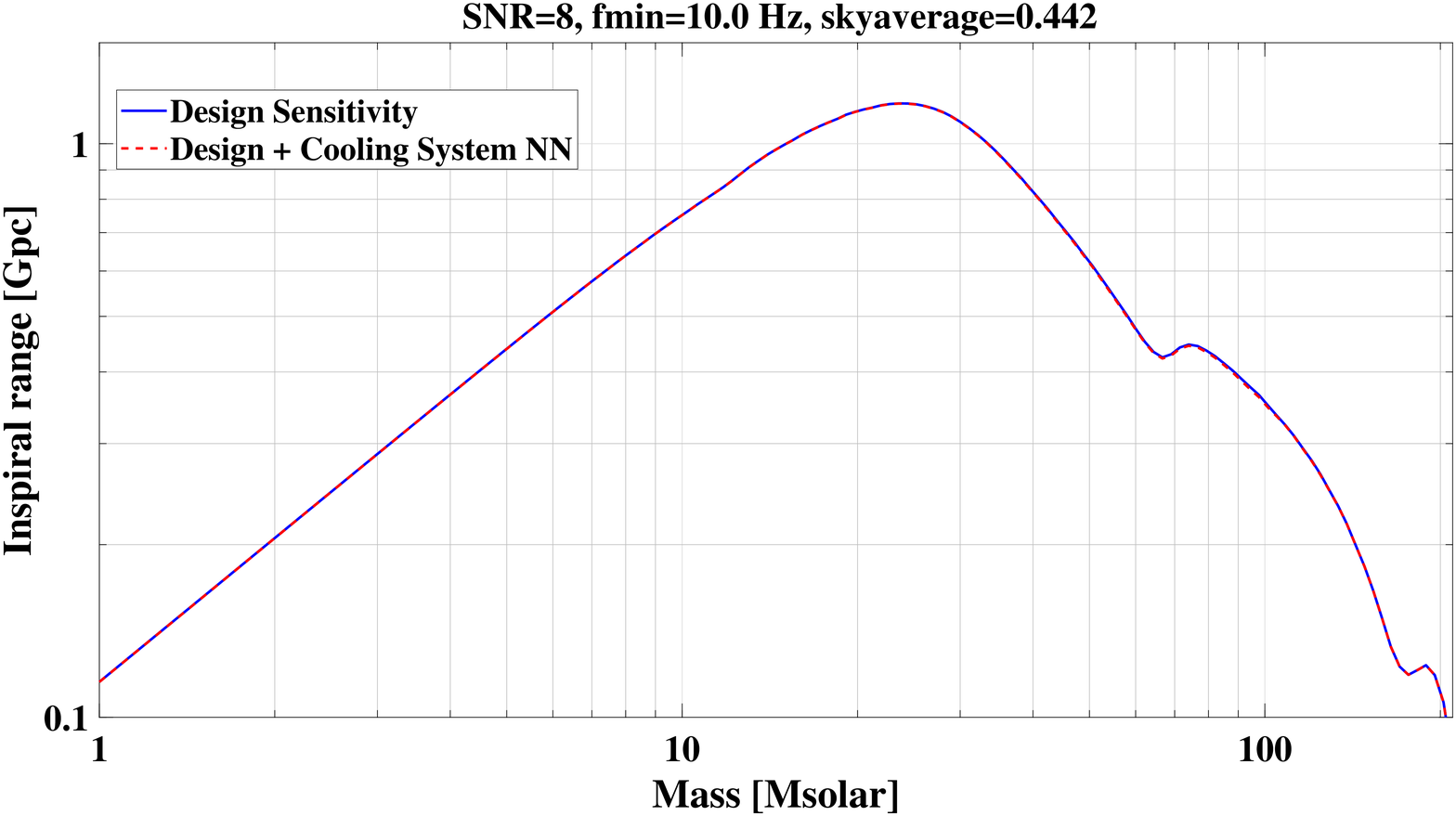}\label{fig:10b}}
	\caption{(a) Comparison of KAGRA design sensitivity and requirement to cooling system Newtonian noise in 1-100 Hz bandwidth. Cooling system NN is below KAGRA design sensitivity but several peaks due cryocooler operation and structural resonances are larger than the requirement. Vibration source of these peaks and contributing component are summarized in \cref{tab:1}. (b) Comparison of KAGRA inspiral range with and without cooling system Newtonian noise. Note that the X-axis represents mass of a component object, assuming equal mass binaries.}
	\label{fig:10}
\end{figure}

\Cref{fig:10a} shows the comparison of Cooling system NN with KAGRA design sensitivities and requirements.
Cooling system NN noise in \cref{fig:10a} is denoted by blue spectra.
NN coupling from the cooling system is lower than design sensitivity but there are several peaks between 16-50 Hz that are larger than the KAGRA design requirements. 
The vibration source and  contributing component for this NN are summarized \cref{tab:1}.
The largest contribution comes from the breadboard except for 21.09 and 21.85 Hz peaks which are due to the shield resonance.
Note that radiation shield contribution is based on the assumption that top and bottom half of shield moves with $u_{\mathrm{8K_{top}}}$ and $u_{\mathrm{8K_{bottom}}}$, respectively. 

\begin{table}[h]
	\caption{Vibration source for cooling system NN peaks that exceed the KAGRA requirement in \cref{fig:10a}. The largest contribution comes from the breadboard except for the peaks marked $"*"$ which are due to radiation shield.}
	\label{tab:1}
	\begin{center}
		\begin{tabular}{|l|l|}
			\hline
			\textbf{Vibration Source} & \textbf{Frequency (Hz)}  \\
			\hline
			Chamber Resonance & 17.2, 21.8\\
			\hline
			Shield Resonance & 16.3, $21.1^*$, $21.8^*$ \\
			\hline
			Cryocooler Operation & 26.8, 35.1, 37.2, 39.3, 41.3,\\
			&43.5, 45.5, 47.6, 46.9\\
			\hline
		\end{tabular}
	\end{center}
\end{table}

We also evaluated the impact of cooling system NN on KAGRAs inspiral range, the results are plotted in \cref{fig:10b}.
The inspiral range calculation was done using MATLAB code from \cite{21} and assumes SNR of 8 and skyaverage of 0.442.
Comparing the inspiral range of KAGRA with (red) and without (blue) cooling system NN, it is clear that this noise does not impact current KAGRA sensitivity.
However, suppressing the cooling system Newtonian noise might be essential in the future, when KAGRA sensitivity is improved  \cite{22}, especially in low frequency. 
As Newtonian noise cannot be shielded against and reduction of vibration will require major design and infrastructure changes; two potential sequential solution are to first remove the breadboard since it has the largest contribution and then to subtract this noise from the interferometer data. 
The standard approach for NN subtraction is to generate Wiener filters using sensor arrays, several studies with this approach have shown promising results for seismic NN cancellation \cite{23,24,25}. 
A similar approach using cryogenic accelerometers \cite{26} should be explored and might be necessary for KAGRA in the future.

\section{Conclusion}
In this paper, we present the estimation of Newtonian Noise coupling form KAGRA cooling system in 1-100 Hz bandwidth. 
First, a simple expression for NN coupling from a point mass-test mass system was derived, this expression was then extended to KAGRA cooling system by breaking each component into mass distribution made of multiple point masses. 
In order to simplify the calculations several considerations were made and the NN coupling for each component was evaluated.

Our calculations show that while cooling system Newtonian Noise is below KAGRA design sensitivity, several peaks above the KAGRA requirements exist in the 16-50 Hz band  due to breadboard and radiation shields Newtonian noise coupling.
The vibration sources for the NN were cryocooler operation and internal resonances of the chamber and radiation shield, making it difficult to attenuate this vibration (and corresponding NN peaks) without significant design and infrastructural changes. 

While this noise does not limit the current detector sensitivity or inspiral range, it might be an issue in the future when KAGRA improves its sensitivity.
Therefore, follow-up studies towards development of Wiener filters using cryogenic accelerometers to subtract cooling system Newtonian noise from KAGRA data are recommended. 
Furthermore, this study also shows that it will be important for the 3rd generation detectors to suppress the vibration of their cooling systems at design and infrastructural level.
\section*{Data Availability}
The data that support the findings of this study are available from the corresponding author upon reasonable request.

\section*{Acknowledgment}
This work was supported by MEXT, JSPS Leading-edge Research Infrastructure Program, JSPS Grant-in-Aid for Specially Promoted Research 26000005, JSPS Grant-in-Aid for Scientific Research on Innovative Areas 2905: JP17H06358, JP17H06361 and JP17H06364, JSPS Core-to-Core Program A. Advanced Research Networks, JSPS Grant-in-Aid for Scientific Research (S) 17H06133, the joint research program of the Institute for Cosmic Ray Research, University of Tokyo, National Research Foundation (NRF) and Computing Infrastructure Project of KISTI-GSDC in Korea, Academia Sinica (AS), AS Grid Center (ASGC) and the Ministry of Science and Technology (MoST) in Taiwan under grants including AS-CDA-105-M06, the LIGO project, and the Virgo project. The authors will like to thank Yuta Michimura for providing the code used to calculate the inspiral range in this paper.

%\section*{References}

\appendix
\section{Vibration Spectra}

\begin{figure}[b]
	\centering
	\includegraphics[width=0.5\textwidth]{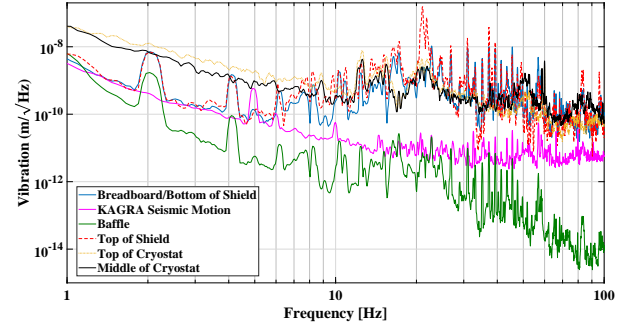}
	\caption{Vibration spectra of various components considered during Newtonian noise estimation.}
	\label{fig:11}
\end{figure}

\Cref{fig:11} shows the vibration spectra of each component and the measurement conditions are mentioned below:
\begin{itemize}
	\item The breadboard vibration ($u_\mathrm{{breadboard}}$) shown above was measured at 12 K using a cryogenic accelerometer \cite{26} in \cite{17}. 
	\item As the breadboard is rigidly bolted to the bottom of inner/8K shield, so the vibration of shield bottom ($u_\mathrm{{8K_{bottom}}}$) will be same as that of breadboard.  
	\item Vibration of top of the inner shield ($u_\mathrm{{80K_{top}}}$) was calculated as the product of breadboard vibration at 12 K and vibration coupling from bottom to top of the shield, measured at room temperature using a commercial accelerometer (\textit{TOKKYOKIKI MG-102S}).
	\item For the outer/80K shield $u_\mathrm{{80K_{bottom}}}=u_\mathrm{{8K_{bottom}}}$ and $u_\mathrm{{80K_{top}}}=u_\mathrm{{8K_{top}}}$ is assumed.
	\item The baffle suspension is rigidly bolted to the breadboard; so it's vibration ($u_\mathrm{{baffle}}$) is the product of breadboard vibration ($u_\mathrm{{breadboard}}$) and suspension transfer function. Since the longitudinal mode of the suspended baffle is 0.84 Hz, the transfer function is: $\frac{0.84^2}{f^2}$ for $f>0.84$ Hz.
	\item Cryostat was divided into three parts, while the top and middle vibrations ($u_\mathrm{{top}}$ and $u_\mathrm{{middle}}$, respectively) were measured by \textit{TOKKYOKIKI MG-102S}, the bottom vibration ($u_\mathrm{{bottom}}$) is same as KAGRA seismic motion and measured using \textit{RION LA-50}.
\end{itemize}	

\begin{thebibliography}{20}
	
	\bibitem{1} Yoichi Aso \textit{et al.}, Interferometer design of the KAGRA gravitational wave detector. 2013 Phys. Rev. D., \textbf{88} 1-15
	
	\bibitem{2} T. Akutsu \textit{et al.}, First cryogenic test operation of underground km-scale gravitational-wave observatory KAGRA. 2019, Class. Quant Grav.., \textbf{36} 165008
	
	\bibitem{3} J.Asai \textit{et al.}, Advanced LIGO. Class. Quantum Grav.., \textbf{32} 074001
	
	\bibitem{4} F. Acernese \textit{et al.}, Advanced Virgo: A second-generation interferometric gravitational wave	detector. 2015, Class. Quantum Grav.., \textbf{32} 024001
	
	\bibitem{5} F Matichard \textit{et al.}, Seismic isolation of Advanced LIGO: Review of strategy, instrumentation and performance. Class. Quantum Grav.., \textbf{32} 185003
	
	\bibitem{6} Ballardin G.; Bracci L.; Braccini S.; Bradaschia C.; Casciano C.; Calamai G; Cavalieri R.; Cecchi R.; Cella G.; Cuoco E. \textit{et al.}, Measurement of the VIRGO superattenuator performance for seismic noise suppression. Review Scientific Instruments., \textbf{72,3643-3652} (2001)
	
	\bibitem{7} Saulson P. R., Terrestrial gravitational noise on a gravitational wave antenna. 1984 Phys. Rev. D., \textbf{30}(3) 732-736
	
	\bibitem{8}  Teviet Creighton, Tumbleweeds and airborne gravitational noise sources for LIGO. 2008, Class. Quantum Grav. \textbf{25}, 125011
	
	\bibitem{9}Pepper K (2007) Newtonian noise simulation and suppression for gravitational-wave interferometers. Technical Note T070192, LIGO, Pasadena, CA, URL \url{https://dcc.ligo.org/LIGO-T070192/public}
	
	\bibitem{10} Donatella Fiorucci, Jan Harms, Matteo Barsuglia, Irene Fiori, and Federico Paolett., Impact of infrasound atmospheric noise on gravity detectors used for astrophysical and geophysical applications. 2018 Phys. Rev. D., \textbf{97},062003
	
	\bibitem{11} Harms, J. Terrestrial gravity fluctuations. \textit{Living Rev Relativ} \textbf{22}, 6 (2019). URL \url{https://doi.org/10.1007/s41114-019-0022-2}
	
	\bibitem{12} M Punturo \textit{et al.}, The Einstein Telescope: a third-generation gravitational wave observatory. 2010, Class. Quantum Grav. \textbf{27}, 194002
	
	\bibitem{13} Reitze, D., Adhikari, R. X., Ballmer, S., Barish, B., Barsotti, L., Billingsley, G., … Zucker, M. (2019). Cosmic Explorer: The U.S. Contribution to Gravitational-Wave Astronomy beyond LIGO. \textit{Bulletin of the AAS, 51(7)}. Retrieved from \url{ https://baas.aas.org/pub/2020n7i035}
	
	\bibitem{14} Francesca Badaracco, Jan Harms, Camilla De Rossi, Irene Fiori, Kouseki Miyo, Taiki Tanaka, Takaaki Yokozawa, Federico Paoletti, Tatsuki Washimi. KAGRA underground environment and lessons for the Einstein Telescope. 2021 Phys. Rev. D., \textbf{104} 042006
	
	\bibitem{15} Somiya K. 2019 Newtonian noise from the underground water. URL \url{http://www-kam2.icrr.u-tokyo.ac.jp/indico/event/3/session/32/contribution/365}
	
	\bibitem{16} C.Tokoku, N. Kimura, S. Koike, \textit{et al.}, Cryogenic system for the interferometric cryogenic gravitational wave telescope, KAGRA - design, fabrication, and performance test. AIP Conference Proceedings \textbf{1573}, 1254 (2014); URL \url{https://doi.org/10.1063/1.4860850}
	
	\bibitem{17} R Bajpai \textit{et al.}, Vibration Analysis of KAGRA Cryostat at Cryogenic Temperature. 2022 \textit{Class. Quantum Grav.} URL \url{https://doi.org/10.1088/1361-6382/ac7cb5}
	
	\bibitem{18} T. Ushiba \textit{et al.}, Cryogenic suspension design for a kilometer-scale gravitational-wave detector. 2021 \textit{Class. Quantum Grav.} \textbf{38} 085013
	
	\bibitem{19} Yoshiyuki Obuchi. Wide Angle Baffle desing review. JGW-G1706474-v1. (JGW Document Server)
	
	\bibitem{20} Simon Zeidler. Recent Activities of the AOS: WAB and NAB. JGW-G1808293-v1. (JGW Document Server)
	
	\bibitem{21} Kentaro Komori. Latest estimated sensitivity of kagra (v201708) *approved*. JGW-T1707038-v9. (JGW Document Server)
	
	\bibitem{22} Y. Michimura \textit{et al.}, Prospects for improving the sensitivity of the cryogenic gravitational wave detector KAGRA. 2020 Phys. Rev. D., \textbf{102}, 022008
	
	\bibitem{23} Cella G. (2000) Off-Line Subtraction of Seismic Newtonian Noise. In: Casciaro B., Fortunato D., Francaviglia M., Masiello A. (eds) Recent Developments in General Relativity. Springer, Milano. pp 495-503. URL \url{https://doi.org/10.1007/978-88-470-2113-6_44}
	
	\bibitem{24} Jennifer C. Driggers, Jan Harms, and Rana X. Adhikari. \textit{et al.}, Subtraction of Newtonian noise using optimized sensor arrays. 2012 Phys. Rev. D., \textbf{86}, 102001
	
	\bibitem{25} Jan Harms and Krishna Venkateswara \textit{et al.}, Newtonian-noise cancellation in large-scale interferometric GW detectors using seismic tiltmeters. 2016 Class. Quantum Grav. \textbf{33} 234001
	
	\bibitem{26} R Bajpai \textit{et al.}, A laser interferometer accelerometer for vibration sensitive cryogenic experiments. 2022 \textit{Meas. Sci. Technol.} \textbf{33} 085902. URL \url{https://doi.org/10.1088/1361-6501/ac6d46}
\end{thebibliography}
\end{document}